\documentclass{ws-procs9x6}
\begin{document}

\def\AJ{{\it Astron. J.} }
\def\ARAA{{\it Annual Rev. of Astron. \& Astrophys.} }
\def\ApJ{{\it Astrophys. J.} }
\def\ApJL{{\it Astrophys. J. Letters} }
\def\ApJS{{\it Astrophys. J. Suppl.} }
\def\ApP{{\it Astropart. Phys.} }
\def\AA{{\it Astron. \& Astroph.} }
\def\AAR{{\it Astron. \& Astroph. Rev.} }
\def\AAL{{\it Astron. \& Astroph. Letters} }
\def\JGR{{\it Journ. of Geophys. Res.}}
\def\JPhG{{\it Journ. of Physics} {\bf G} }
\def\PhFl{{\it Phys. of Fluids} }
\def\PR{{\it Phys. Rev.} }
\def\PRD{{\it Phys. Rev.} {\bf D} }
\def\PRL{{\it Phys. Rev. Letters} }
\def\PLB{{\it Phys. Lett. B}}
\def\Nature{{\it Nature} }
\def\NewAR{{\it New Astron. Rev.}}
\def\MNRAS{{\it Month. Not. Roy. Astr. Soc.} }
\def\ZA{{\it Zeitschr. f{\"u}r Astrophys.} }
\def\ZFN{{\it Zeitschr. f{\"u}r Naturforsch.} }
\def\etal{{\it et al.}}
%
\def\simle{\lower 2pt \hbox {$\buildrel < \over {\scriptstyle \sim }$}}
\def\simge{\lower 2pt \hbox {$\buildrel > \over {\scriptstyle \sim }$}}


\title{THE NATURE OF DARK MATTER}

\author{Peter L. Biermann\footnote{E-mail:plbiermann@mpifr-bonn.mpg.de}$^{1,2,3}$,
and  
Faustin Munyaneza\footnote{E-mail:munyanez@mpifr-bonn.mpg.de}\footnote{Humboldt Fellow}$^1$}
\address{
$^1$Max-Planck Institute for Radioastronomy, Bonn, Germany\\
$^2$Department of Physics and Astronomy,
University of Bonn,  Germany, \\
$^3$Department of Physics and Astronomy, University of Alabama,
Tuscaloosa, AL, USA
}

\begin{abstract}
Dark matter has been recognized as an essential part of matter for over 70
years now, and many suggestions have been made, what it could be.  Most of 
these ideas have centered on Cold Dark Matter, particles that are expected 
in extensions of standard particle physics, such as supersymmetry.  Here
we explore the concept that dark matter is sterile neutrinos, a concept
that is commonly referred to as Warm Dark Matter.  Such particles 
have keV masses, and decay over a very long time, much longer than the 
Hubble time.  In their decay they produce X-ray photons which modify the
ionization balance in the early universe, increasing the fraction of
molecular Hydrogen, and thus help early star formation.  Sterile neutrinos
may also help to understand the baryon-asymmetry, the pulsar kicks, the
early growth of black holes, the minimum mass of dwarf spheroidal
galaxies, as well as the shape of dark matter halos.  As soon as all these
tests have been quantitative in its various parameters, we may focus on
the creation mechanism of these particles, and could predict the strength
of the sharp X-ray emission line,
expected from any large dark matter assembly.  A measurement of this X-ray
emission line would be definitive proof for the existence of may be called
weakly interacting neutrinos, or WINs.

\end{abstract}

\bodymatter

\section{Dark Matter: Introduction}

Since the pioneering works of Oort \cite{oort32} and Zwicky \cite{zwicky33,zwicky37},
we know that there is
dark matter in the
universe, matter that interacts gravitationally, but not measureably in
any     other way.  Oort argued about the motion and density of stars
perpendicular to
the Galactic plane, and in this case, Oort's original hunch proved to be
correct, the missing matter turned out to be low luminosity stars.  Zwicky
argued about the motions and densities of galaxies in clusters of
galaxies, and to this day clusters of galaxies are prime arguments to
determine dark matter, and its properties.

Based on the microwave back ground fluctuations\cite{spergel06} today we know that the universe is flat geometrically, i.e. the
sum of the angles in a
cosmic triangle is always 180 degrees, provided we do not pass too close
to a
black hole.   This finding can be translated into stating that the sum of
the mass and energy contributions to the critical density of the universe
add up to unity, with about 0.04 in baryonic matter, about 0.20 in dark
matter, and the rest in dark energy;  we note that there is no consensus
even where to find all the baryonic matter, but a good guess is warm to
hot gas, such as found in groups and clusters of galaxies, and around
early Hubble type galaxies.

There are many speculations of what dark matter is; we have three
constraints:

1)  It interacts almost exclusively by gravitation, and not measurably in
any other way;  2)  It does not participate in the nuclear reactions in
the early universe; 3)  It must be able to clump, to help form galaxies,
and later clusters of galaxies, and the large scale structure.

Obviously, various extensions in particle physics theory, such as
supersymmetry, all provide candidates, like the lightest supersymmetric
particle.

Here we focus on the concept that it may be a ``sterile neutrino", a
right-handed neutrino, that interacts only weakly with other neutrinos,
and otherwise only gravitationally.  Such particles were  first proposed 
 by
Pontecorvo \cite{pontecorvo70} and later by  Olive \& Turner  \cite{olive82}.
Sterile neutrinos  were further proposed
 as
 dark matter candidates \cite{dodelson94}.
  It was then shown how
oscillations of normal neutrinos to sterile neutrinos could help explain
the very large rectilinear velocities of some pulsars \cite{kusenkosegre97}.

Observationally the evidence comes from a variety of arguments:
i)  Dark matter in a halo like distribution is required to explain the
stability of spiral galaxy disks \cite{ostriker73,ostriker74};
 ii) the
flat rotation curves of galaxies \cite{rubin80}); and iii)  the
containment of hot gas in early Hubble type galaxies \cite{biermann82}.
 Dark matter is required to explain iv)  the structure of
clusters of galaxies\cite{ensslin97}; v)  structure formation,
and  the flat geometry of the universe \cite{spergel03,spergel06}.
We refer the reader to a  recent review on dark matter \cite{bertone04}.

Therefore after more than 70 years we still face the question: ``What is
dark matter?"

\section{Proposal}

The existing proposals to explain dark matter mostly focus on very massive
particles \cite{bertone04}, such as the
lightest
supersymmetric particle; all the experimental searches are sensitive for
masses above GeV, usually far above such an energy.  In the normal
approach to structure formation, this implies a spectrum of dark matter
clumps extending far down to globular cluster masses and below.  It has
been a difficulty for some time that there is no evidence for a large
number of such entities near our Galaxy.  The halo is clumpy in stars, but
not so extremely clumpy.  If, however, the mass of the dark matter
particle were in the keV range, then the lowest mass clumps would be large
enough to explain this lack.  However, in this case the first star
formation would be so extremely delayed \cite{yoshida03}
 that there
would be no explanation of the early reionization of the universe, between
redshifts 11 and 6, as we now know for sure \cite{spergel03,spergel06}.
 Therefore, the conundrum remained.

Here we explore the concept that the dark matter is indeed of a mass in
the
keV range, but can decay, and so produce in its decay a photon, which
ionizes, so modifies the abundance of molecular Hydrogen, and so allows
star formation to proceed early \cite{biermannkusenko06,jarek06}.
 The specific model we explore is of ``sterile
neutrinos", right handed neutrinos, which interact only with normal,
left-handed neutrinos, and with gravity.  Such particles are commonly
referred to as ``Warm Dark Matter", as opposed to ``Cold Dark Matter",
those very massive particles. For most aspects of cosmology warm dark
matter and cold dark matter predict the same; only at the small scales are
they significantly different, and of course in their decay.

The mass range we explore is approximately 2 - 25 keV.  These sterile
neutrinos decay, with a very long lifetime, and in a first channel give
three normal neutrinos, and in the second channel, a two-body decay, give
a photon and a normal neutrino. The energy of this photon is almost
exactly half the mass of the initial sterile neutrino.

What is important is to understand that such particles are not produced
from any process in thermal equilibrium, and so their initial phase space
distribution is far from thermal; all the current models for their
distribution suggest that their momenta are sub-thermal. The measure of
how much they are sub-thermal modifies the precise relationship between the
dark matter particle mass and the minimum clump mass, which should be
visible in the smallest pristine galaxies.

This also entails, that as Fermions they require a Fermi-Dirac
distribution, as
being far from equilibrium, this distribution implies a chemical
potential.

Recent work by many others \cite{abazajian01, abazajian05, abazajiankoushiapas06,
abazajian06a,abazajian06b, boyarsky06a,boyarsky06b,boyarsky06c, dolgovhansen02, 
asakashapo05, asaka05,
asaka06}
  has shown that these sterile neutrinos
can be produced in the right amount to explain dark matter, could explain
the baryon asymmetry \cite{akhmedov98}, explain the lack of power on
small scales (as noted above), and could explain the dark matter
distribution in galaxies \cite{belokurov06,fellhauer06,gilmore06}.

\subsection{Our recent work}

Pulsars are observed to reach linear space velocities of up to over 1000
km/s, and there are not many options how to explain this; one possibility
is to do this through magnetic fields which become important in the
explosion\cite{bisnovatyi70, bisnovatyi93,bisnovatyi95,ardelian05,moiseenko05}.
  Another possibility is to do this through a
conversion of active neutrinos which scatter with a mean free path of
about ten cm, into sterile neutrinos, which no longer scatter.  If this
conversion produces a spatial and directional correlation between the
sterile neutrinos and the structure of the highly magnetic and rotating
core of the exploding star, then a small part of the momentum of the
neutrinos can give an asymmetric momentum to the budding neutron star,
ejecting it at a high velocity\cite{kusenko04}.  This then could explain
such features as the guitar nebula, the bow shock around a high velocity
pulsar.  This latter model in one approximation requires a sterile
neutrino in the mass range 2 to 20 keV. It is remarkable that this
neutrino model requires magnetic fields in the upper range of the
strengths predicted by the magneto-rotational mechanism to explode massive
stars as supernovae.

It was also shown from SDSS data, that some quasars have
supermassive black holes already at redshift 6.41, so 800 million years
after the big
bang \cite{fan01,willot03}.  We now know, that this is exactly when galaxies grow the fastest,
from 500 to 900 million years after the big bang \cite{bouwens06,iye06}
  Baryonic accretion has trouble feeding a normal black hole to
this high mass, $ 3 \; 10^{9}$ solar masses so early after the big bang,
if the growth were to start with stellar mass black holes \cite{wangbiermann98}.
  So either the first black holes are around $10^{4}$ to
$10^{6}$ solar masses, and there is not much evidence for this, or the
early black holes grow from dark 
matter\cite{munyanezabiermann05,munyanezabiermann06,munyanezabiermann06a},
 until they reach the critical minimum mass to be able to grow very
fast and further from baryonic
matter, which implies this mass range, $10^{4}$ to $10^{6}$ solar
masses.  This model in the isothermal approximation for galaxy structure
implies a
sterile neutrino in the mass range between 12 and 450 keV.

When Biermann and Kusenko met at  Aspen meeting September 2005,
it became apparent, that these two speculative approaches overlap, and so
it seemed worth to pursue them further.

As noted above, structure formation arguments lead to an over-prediction in
power at small scales in the dark matter distribution in the case of cold
dark matter, and any attempt to solve this with warm dark matter delayed
star formation unacceptably. We convinced ourselves that this was the key
problem in reconciling warm dark matter (keV particles) with the
requirements of large scale structure and reionization.
 We then showed that the decay of the sterile neutrino could
increase the ionization, sufficiently to enhance the formation of
molecular hydrogen, which in turn can provide catastrophic cooling early
enough to allow star formation as early as required \cite{biermannkusenko06,jarek06}.
In our first simple calculation this happens at redshift 80.  More refined
calculations confirm, that the decay of sterile neutrinos helps increase
the fraction of molecular Hydrogen, and so help star formation, as long as
this is at redshifts larger than about 20 \cite{mapelliferrara05,mapelli06,ripamonti06}.

\section{The tests}

\subsection{Primordial magnetic fields }

In the decay a photon is produced, and this photon ionizes Hydrogen: at
the first ionization an energetic electron is produced, which then ionizes
much
further, enhancing the rate of ionization by a factor of about 100.  In
the case, however, that there are primordial magnetic fields, this
energetic electron could be caught up in wave-particle interaction, and
gain energy rather than lose energy.  As the cross section for ionization
decreases with energy, the entire additional ionization by a factor of
order 100 would be lost in this case, and so there basically would be no
measurable effect from the dark matter decay.  This gives a limit for the
strength of the primordial magnetic field, given various models for the
irregularity spectrum of the field:  In all reasonable models this limit
is of order a few to a few tens of picoGauss, recalibrated to today.
Recent simulations matched to the magnetic field data
of clusters and superclusters, give even more stringent limits, of
picoGauss or less \cite{dolag05}.

It follows that primordial magnetic fields can not disturb the early
ionization from the energetic photons, as a result of dark matter
decay.  It then also follows that the contribution of early magnetic
fields from magnetic monopoles, or any other primordial mechanism, is
correspondingly weak\cite{wick03}.

Stars at all masses are clearly able to produce magnetic 
fields \cite{biermannl50, biermannl51, silklanger06},
but the evolution and consequent dispersal are fastest for the massive
stars, almost certainly the first stars.  As the magnetic fields may help
to drive the wind of these massive stars \cite{seemannbiermann97},
  then
the wind is just weakly super-Alfv\'enic, with Alfv\'enic Machnumbers of
order a few.  This implies that the massive stars and their winds already
before the final supernova explosion may provide a magnetic field which is
at order 10 percent equipartition of the environment; this magnetic field
is highly structured.  However, even these highly structured magnetic
fields will also allow the first cosmic rays to be produced, and
distributed, again with about 10 percent of equipartition of the
environment.

However, the large scale structure and coherence of the cosmic magnetic
fields clearly remain an unsolved problem \cite{kulsrud99,biermanngalea03,biermannkronberg04}.
 
Therefore the first massive stars are critical for the early evolution of
the universe:  In addition to reionization, magnetic fields and cosmic
rays, they provide the first heavy elements.  These heavy elements allow
in turn
dust formation, which can be quite rapid (as seen, e.g., in SN 1987A,
already
just years after the explosion \cite{biermann90}.  This then
enhances the cooling in the dusty regions, allowing the next generation of
stars to form much faster.

In combination everywhere one first massive star is formed, we can
envisage a runaway in further star formation in its environment.

\subsection{Galaxies}

Galaxies merge, and simulations demonstrate that the inner dark matter
distribution attains a power law in density, and a corresponding power law
tail in the momentum distribution \cite{nfw97,moore99,klypin02}:
 Here the central density distribution as a
result of the merger is a divergent power law, as a result of energy
flowing outwards and mass flowing inwards, rather akin to accretion 
disks \cite{lust52,lustschluter55}
 where angular momentum flows outwards and mass also flows
inwards; in fact also in galaxy mergers angular momentum needs to be
redistributed outwards as such mergers are almost never central \cite{toomre72}.
 This then leads to a local escape velocity converging with
$r$ to zero also towards zero, and so for fermions the Pauli limit is
reached, giving rise to a cap in density, and so a dark matter star or a fermion ball is
born \cite{munyanezabiermann05,munyanezabiermann06};
 this dark matter star can grow
further by dark matter accretion. The physics of fermion balls at galactic centers
has been  studied in a series of 
papers\cite{viollier94,tv98,mtv98,mtv99,bmv99,mv02,bmtv02}.
  For realistic models an integral over a
temperature distribution is required, and a boundary condition has to be
used to represent the surface of the dark matter star both in real space
as in momentum space.  This then allows the mass of this dark matter star
to increase; such models resemble in their quantum statistics white dwarf
stars or neutron stars; the Pauli pressure upholds the star.  For fermions
in the keV range the mass of the dark matter star has a mass range of a
few thousand to a few million solar masses.

The first stellar black hole can then enter this configuration and eat the
dark matter star from inside, taking particles from the low angular
momentum phase space.  With phase space continuously refilled through the
turmoil of the galaxy merger in its abating stages, or in the next merger,
the eating of the dark matter star from inside ends only when all the dark
matter star has been eaten up.

Given a good description of the dark matter star boundary conditions in
real and in momentum phase space \cite{munyanezabiermann05,munyanezabiermann06},
  and an observation of the stellar velocity dispersion close to the
final black hole, but outside its immediate radial range of influence, we
should be able to determine a limit to the dark matter particle mass.  If
the entire black hole in the Galactic Center has grown from dark matter
alone, then we obtain a real number.

This concept suggests that it might be worthwhile to consider the smallest
of all black holes in galactic centers.  In a plot of black hole mass
versus central stellar velocity dispersion $\sigma$  there is a clump above the relation  $M_{BH} \; \sim
\; \sigma^{4}$, at low black hole masses \cite{barth05,greene06}, suggesting that perhaps 
we reach
a limiting relationship with a flatter slope for all those black holes
which grow only from dark matter \cite{munyanezabiermann06}; for a simple isothermal approach
 this
flatter slope is found to be 3/2.

\subsection{Dwarf spheroidal galaxies}

All detected dwarf spheroidal galaxies
fit a simple limiting relationship of a common dark matter mass of $5 \,
10^7$ solar masses \cite{gilmore06}, suggesting that this is perhaps the smallest dark
matter clump mass in the initial cosmological dark matter clump
spectrum.  This clump mass is of course a lower limit to the true original
mass of the pristine dwarf spheroidal galaxy.  Given a physical concept
for the production of the dark matter particles in the early universe, we
would have their initial momentum, probably subthermal, and so the
connection between the dark matter particle mass and minimum clump mass is
modified.  This is very strong support for the Warm Dark Matter concept.

One intriguing aspect of dwarf spheroidal galaxies is that almost all of
them show the effect of tidal distortion in their outer regions, and at
least one of them has been distended to two, perhaps even three
circumferential rings around our Galaxy \cite{belokurov06,fellhauer06}.
 To extend so far around our Galaxy must have taken many
orbits, and so a some fraction of the age of our Galaxy.  The simple
observation that these streamers still exist separately, and can be
distinguished in the sky, after many rotations around our Galaxy, implies
that the dark matter halo is extremely smooth, and also nearly spherically
symmetric. Given that the stellar halo is quite clumpy this implies once
more that the dark matter is much more massive than the baryonic matter in
our halo.

\subsection{Lyman alpha forest}

In the early structure formation the large number of linear perturbations
in density do not lead to galaxies, but just too small enhancements of
Hydrogen density, visible in absorption against a background quasar.  This
so-called Lyman alpha forest tests the section of the perturbation scales
which is linear and so much easier to understand, and it should in
principle allow a test for the smallest clumps \cite{viel06}.
 Unfortunately, systematics make this test still difficult, and
with the expected sub-thermal phase space distribution of the dark matter
particles we may lack yet the sensitivity to determine the mass of the
smallest clumps.

\subsection{The X-ray test}

When the sterile neutrinos decay, they give off a photon with almost
exactly half their mass in energy.  Our nearby dwarf spheroidal galaxies,
our own inner Galaxy, nearby massive galaxies like M31, the next clusters
of galaxies like the Virgo cluster, and other clusters further away, all
should show a sharp X-ray emission line \cite{abazajian01,riemer06a,riemer06b,watson06}.

The universal X-ray background should show such a sharp emission line as a
wedge, integrating to high redshift.

With major effort this line or wedge  be detectable with the current
Japanese, American or European X-ray satellites: Large field high spectral
resolution spectroscopy is required.

\section{Outlook}

The potential of these right handed neutrinos is impressive, but in all
cases we have argued, there is a way out, in each case there is an
alternative way to interpret the data set.  E.g., for the pulsar kick with
the help of neutrinos strong magnetic fields are required, but the MHD
simulations suggest that perhaps magnetic fields can do it by themselves,
even without the weakly interacting neutrinos
\cite{bisnovatyi70,ardelian05,moiseenko05}.
The dwarf spheroidal
galaxies can in some models be explained without any dark matter 
at all \cite{metz06,sohn06}.
 The early growth of black holes can
also be fueled by other black holes, as long as here are enough in number
and their angular momentum can be removed.  So many alternatives may
replace the sterile neutrino concept.

However, the right handed, sterile neutrinos weakly interacting with the
normal left handed neutrinos provide a unifying simple hypothesis, which
offers a unique explanation of a large number of phenomena, so by Occam's
razor, it seems quite convincing at present\cite{kusenko06}.  So, given
what sterile neutrinos may effect, we may have to call them Weakly
Interacting Neutrinos, or soon WINs.

\section{Acknowledgements}
The authors wish to thank first and foremost Alex Kusenko, an
indefatigable partner in all explorations of warm dark matter; he played
a key role in working out  the science reported here.
The authors would  also like to acknowledge fruitful discussions with Kevork
Abazajian, Gennadi Bisnovatyi-Kogan, Gerry Gilmore, Phil Kronberg, Pavel
Kroupa, Sergei Moiseenko, Biman Nath, Mikhail Shaposhnikov, Simon Vidrih,
Tomaz Zwitter, and many others.

Support for PLB is coming from the AUGER membership and theory grant 05
CU 5PD 1/2 via DESY/BMBF. Support for FM is coming from the Humboldt Foundation.

\end{document}